\documentclass[a4paper,12pt]{spieman}  
\usepackage{amsmath,amsfonts,amssymb}
\usepackage{graphicx}
\usepackage{setspace}
\usepackage{tocloft}
\usepackage{lineno}
\usepackage{orcidlink}
\graphicspath{ {pics/} }

\title{Design, Fabrication and Characterization of the Thermal Filter Assembly on the Solar Ultraviolet Imaging Telescope (SUIT) on-board Aditya- L1}
\author[a,b]{Janmejoy Sarkar \orcidlink{0000-0002-8560-318X}} 
\author[a,d,h]{Avyarthana Ghosh \orcidlink{0000-0002-7184-8004}}
\author[c,a]{Sreejith Padinhatteeri \orcidlink{0000-0002-7276-4670}}
\author[a]{Ravi Kesharwani \orcidlink{0009-0002-2528-5738}}
\author[a,d]{Ramaprakash A.N. \orcidlink{0000-0001-5707-4965}}
\author[a,d]{Durgesh Tripathi \orcidlink{0000-0003-1689-6254}}

\author[e]{Bhargava Ram B.S. \orcidlink{0000-0001-7634-1790}}
\author[e]{R. Venkateshwaran}
\author[g]{Ketan Patel}
\author[a]{Melvin James}
\author[d]{Mintu Karmakar}
\author[a]{Akshay Kulkarni}
\author[a]{Deepa Modi}
\author[a]{Chaitanya Rajarshi}
\author[e]{Girish M. Gouda} 
\author[a]{Aafaque R. Khan \orcidlink{0000-0002-1244-0295}}
\author[f]{Abhijit Adoni}
\author[f]{Sajjade F. Mustafa}
\author[a]{Pravin Khodade}
\author[a]{Abhay Kohok}

\affil[a]{Inter-University Centre for Astronomy and Astrophysics, Post Bag 4, Ganeshkhind, Pune - 411007, Maharashtra, India}
\affil[b]{Department of Physics, Tezpur University, Napaam, Tezpur-784028, Assam, India}
\affil[c]{Manipal Centre for Natural Sciences, Manipal Academy of Higher Education, Karnataka, Manipal- 576104, India}
\affil[d]{Center of Excellence in Space Sciences India, Indian Institute of Science Education and Research Kolkata, Mohanpur 741246, West Bengal, India}
\affil[e]{Laboratory for Electro-Optics Systems (LEOS), ISRO, $1^{st}$ Cross, $1^{st}$ Phase, Peenya, Bengaluru- 560058, Karnataka}
\affil[f]{U R Rao Satellite Centre,  Old Airport Road Vimanapura Post, Bengaluru - 560017, Karnataka, India}
\affil[g]{Luma Optics Pvt. Ltd., Mumbai, India }
\affil[h]{Now at EDIS, TCS Research, India}

\newcommand{\suit}{{\it SUIT~}}
\newcommand{\degree}{{$^{\circ}$}}
\newcommand{\js}[1]{{\color{black} {#1}}}

\cftpagenumbersoff{figure}
\cftpagenumbersoff{table} 

\begin{document} 
\maketitle

\begin{abstract}
The Solar Ultraviolet Imaging Telescope (\suit) observes the Sun in the near-ultraviolet regime on board the Aditya-L1 satellite, India's dedicated mission to study the Sun. \suit will image the Sun in the wavelength range of 200{--}400~nm using 11 science \js{bandpasses} with varying spectral bandwidths between 0.1{--}58~nm.
Within this range, the Sun provides huge incoming solar flux to the telescope that also varies by a factor of $\approx$ 20 from the lower end to the upper end of the wavelength band of interest. Thermal Filter Assembly (TFA) is an optical component at the SUIT entrance aperture, directly facing the Sun. The TFA is used to control the heat load entering the telescope cavity and also to reduce the signal reaching the \suit camera system and the charge-coupled device (CCD) sensor, which is limited in full-well capacity and the linear operational regime. The TFA is designed to allow only {0.1{--}0.45\%} of the incoming flux to pass within 200{--}400~nm. The choice of materials for substrate and coating for the filter poses several challenges in terms of contamination, corrosion/ oxidation and durability during the manufacturing process. Additionally, long-term exposure to harsh space environments and the formation of pinholes are other concerns. Direct exposure to the sun leads to a strong temperature gradient along the thickness of the filter. The design and assembly of the TFA are performed to avoid any thermo-elastic stress affecting optical performance. Different levels of qualification tests and the operation of \suit in orbit for more than 14 months have confirmed the perfect working of the TFA. To the best of our knowledge, the design, development, and testing of such a rejection filter is the first of its kind for space telescopes in the near ultraviolet range.
\end{abstract}


{\noindent \footnotesize\textbf{Corresponding Author: Sreejith Padinhatteeri}  \linkable{sreejith.p@manipal.edu} }

\begin{spacing}{2}   

\section{Introduction}\label{sec:intro}
\begin{figure}
	\centering
	\includegraphics[width=0.9\linewidth]{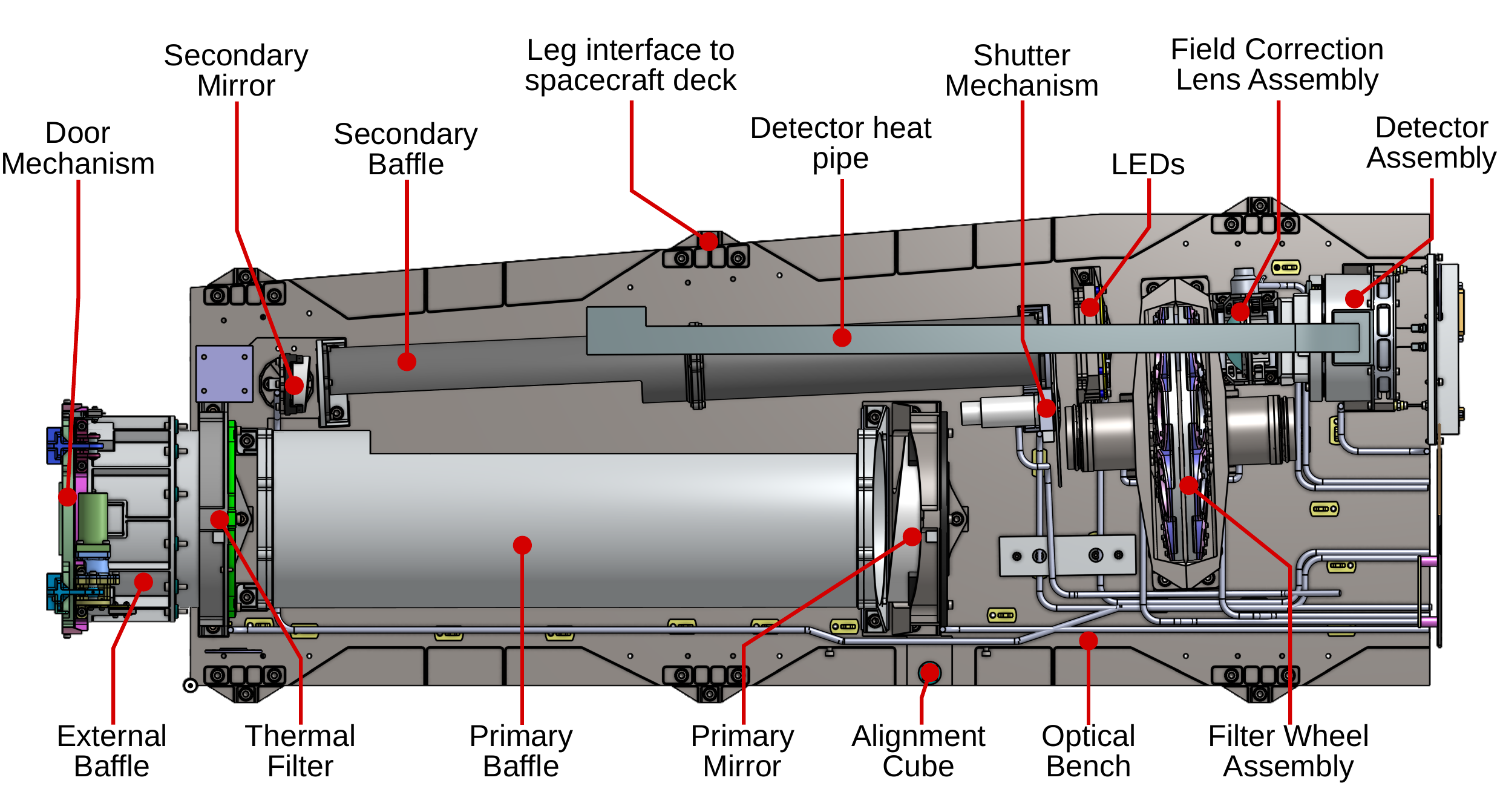}
	\caption{Schematic diagram of \suit onboard Aditya-L1.}
	\label{fig:payload}
\end{figure}

The Solar Ultraviolet Imaging Telescope \cite{GhoCK_2016, TriRK_2017, Tripathi2025_SUITMain} is one of the payloads on board the Aditya-L1 \cite{TriC_2023, SeeM_2017} space solar observatory of the Indian Space Research Organization (ISRO). SUIT is an off-axis Ritchey–Chrétien telescope observing the Sun in the 200 {--} 400 nm regime with 8 narrow and 3 broad bandpasses. To facilitate this, SUIT uses specific combinations of 16 dichroic filters mounted on two filter wheels with 8 filters each. The imaging is performed by a back-thinned, back-illuminated, UV enhanced CCD detector.
Scientific objectives of SUIT encompass a variety of subjects, which can be summarized to three major points,  $viz.,$  ``1. Studying the photospheric and chromospheric activities and it's coupling between the layers of solar atmosphere, 2. energetics and evolution of flares and prominence eruptions, 3. spatially resolved imaging of the Sun in near ultraviolet (NUV) to study the UV-variability that could affect the earth climate and space weather" \cite{Tripathi2025_SUITMain}. A schematic diagram of the telescope is shown in Fig.~\ref{fig:payload}. 

The biggest challenge for solar telescopes is to prevent detector saturation due to the high radiance in the visible range. \js{Mitigating this with fast exposures would demand extremely short exposure times, which is tough to achieve with high accuracy.}

\js{
To mitigate this, successive generations of solar telescopes have demonstrated the use of an entrance aperture heat rejection filter.

Extreme ultraviolet telescopes such as the Extreme-Ultraviolet Imaging Telescope (EIT) \cite{DelAB_1995} on-board the SOlar Heliospheric Observatory (SOHO) \cite{DomFP_1995}, and the Extreme-Ultraviolet Imaging Spectrometer \cite{CulHJ_2007} on-board Hinode \cite{KosMS_2007} used entrance aperture heat rejection filters.
The Transition Region And Coronal Explorer (TRACE) \cite{HanAK_1999} on board Hinode used an additional thin film aluminum filter, along with the heat rejection filter, to compensate for light leaking through pinholes and other failures of the filters.

Further improvements in this regard have been attained for later instruments like the Atmospheric Imaging Assembly (AIA) \cite{BoeWT_2011, BoeEL_2012, LemTA_2012} and the EUV Variability Experiment (EVE) \cite{WooEH_2012}, onboard the Solar Dynamics Observatory (SDO) \cite{PesTC_2012}.
AIA uses a combination of thin metal sheets at the entrance aperture, along with focal plane aluminum and zirconium films to cut down visible and infrared light by a factor of $10^{-11}$ \cite{CheCC_2009}.
The Multiple EUV Grating Spectrographs- A (MEGS-A) on EVE uses thin foil filters to allow light between 5{--}37~nm into the instrument. Multiple EUV Grating Spectrographs-B (MEGS-B), on the other hand, uses multiple gratings to allow light between 35{--}105~nm to fall on the detector \cite{CroWT_2004, CroWT_2007}.

Alternative to entrance aperture filters, the Solar Ultraviolet Measurements of Emitted Radiation \cite{WilCM_1995}, on-board the SoHO, used a combination of a heat-rejecting mirror, baffle system, light traps, and an aperture stop at the entrance.
The Solar Optical Telescope (SOT) \cite{TsuIK_2008} onboard the Hinode, is equipped with an IR-rejection filter at the entrance of the collimating lens unit (CLU), in addition to a heat dump mirror (HDM) at the primary focus to get rid of the unwanted infrared solar radiation \cite{SueIK_2017}.

The Full Sun Imager \cite[and references therein]{RocAB_2020} of the Extreme Ultraviolet Imager on board the Solar Orbiter \cite{HalPT_2010, RocAB_2020} uses a 150 nm thick aluminum filter of size 20~mm $\times$ 20~mm, arranged appropriately in the ray path for achieving the desired rejection of visible and infrared radiation. 
The second payload on board the High Resolution Imager (HRI) \cite{HalMM_2015}, has an aluminum foil filter between the entrance pupil and the primary mirror. This reduces excessive heat input on the mirror and allows efficient visible light rejection.

In addition to heat rejection, some instruments develop filters that allow the transmission of specific scientifically significant bands of light.
The HRI Lyman-$\alpha$ (HRI$_{Lya}$) unit uses a broad-band interference filter (Type 122 NB-40D of Pelham Research Optical LLC) at the entrance of the telescope, to isolate the 121.6~nm spectral line from the rest of the solar spectrum \cite{SchJS_2011}. 
The XUV Photometer System (XPS), a part of the Solar Extreme ultraviolet Experiment (SEE) onboard the TIMED Satellite \cite[and the references therein]{WooW_2004}, deployed fused silica windows on the filter wheel mechanism to allow a wavelength range of 0.1{--}34~nm.
}

In this context, it is imperative to mention the \js{Sunrise Filter Imager (SuFI) \cite{gandorfer_filter_2011}} on board the balloon-borne Sunrise observatory \cite{SolBD_2010, SolR_2017} and the Interface Region Imaging Spectrometer (IRIS)\cite{DePTL_2014}, which operate in the near ultraviolet band like SUIT.
SuFI used imaging filters with additional dielectric coatings in pairs, to suppress out-of-band radiation of longer wavelengths. The IRIS telescope also uses similar dielectric coatings on all components in the optical path to achieve high suppression of visible and infrared \cite{WulJT_2012, PodCG_2012, ChePB_2014}.

SUIT uses an entrance aperture Thermal Filter (TF) which is depicted in Figure \ref{fig:payload}. In combination with SUIT bandpass selection filters, this helps to suppress out-of-band radiation and accommodate the NUV radiation within the dynamic range of the SUIT detector. Additionally, it regulates the heat load inside the optical cavity by maintaining the SUIT optical bench temperature within operational limits of (20$\pm$1$^{\circ}$C).

In this paper, we describe the significance and the role of the Thermal Filter Assembly (TFA) in achieving the desired performance of the SUIT payload.
We discuss the design considerations in \S \ref{sec:design} and present the fabrication methods in \S \ref{sec:fabrication}. The details of space and launch environmental qualification tests performed are discussed in \S \ref{sec:qualification}. The optical testing and characterization of the subassembly is discussed in \S \ref{sec:testing}.

\begin{figure*}
	\centering
	\includegraphics[width=0.8\textwidth]{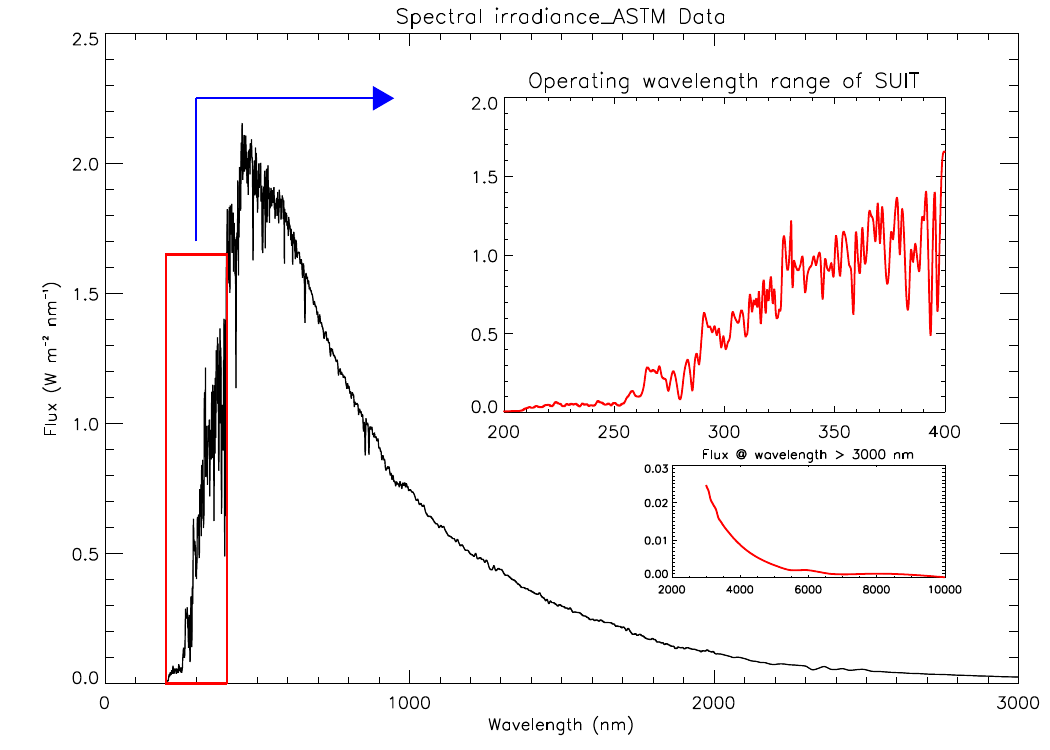}
	\caption{The solar spectral irradiance within 0{--}10000~nm as obtained from the ASTM. The highlighted red box marks the operational range of SUIT. Upper inset: expanded view for the same. Lower inset: the solar spectrum $>$ 3000~nm \cite{th_filt_spie}.}\label{spectra}
\end{figure*}

\begin{table}
	\caption{The identifiers for the SFs (column 1) and CFs (column 2) along with their region of observation, central wavelengths, theoretical FWHM and passbands (PBs) are noted here. Here, NB and BB stand for Narrowband and Broadband filters, respectively.}\label{scfil}
	\begin{tabular}{@{}llllll@{}}
		\hline
		Identification  & Filter central    & FWHM      & Complementary & Passband      & Remarks   \\
		&wavelength         &           & filter        &               &  \\
		&(nm)               & (nm)      &               &(nm)           &  \\     
		\hline
		NB01             &214                & 11.55     &  BB01          &200.00-235.80  & Continuum \\
		NB02             &276.7             & 0.4       &  BP02          &275.94-277.46  & Continuum \\
		NB03             &279.6             & 0.4       &  BP02          &277.60-281.61  & Mg II k   \\ 
		NB04             &280.3             & 0.4       &  BP02          &278.62-281.98  & Mg II h   \\ 
		NB05             & 283.2            & 0.4       &  BP02          &282.25-284.15  & Continuum \\
		NB06             &300                & 1.0       &  BP03          &297.45-302.55  & Continuum \\
		NB07             &388                & 1.0       &  BP03          &384.35-391.65  & Continuum \\ 
		NB08             &396.85             & 0.1       &  NB08          &395.80-397.90  & Ca II h   \\ 
		BB01             &200-242            & 42.0      &  BB01          &200-242        & Continuum \\
		BB02             &242-300            & 58.0      &  BP04          &242-306        & Continuum \\
		BB03             &320-360            & 40.0      &  BP04          &320-360        & Continuum\\
		\hline
	\end{tabular}
\end{table}

\section{Design Parameters \& Constraints} \label{sec:design}
The challenges in the design and fabrication of the thermal filter are posed by its low weight requirement and tolerance to the space environment at the Sun-Earth Lagrange 1 point. The substrate and coating materials for the thermal filter should be able to endure thermal shocks, high energy radiation, and huge temperature gradients along the optical axis. The filter coating should also be resistant to corrosion, oxidation, or any other degradation like formation of pinholes while handling or in space.

The following design considerations are kept in mind while designing the thermal filter- 1. The detector performance and response linearity within the operational wavelength range of SUIT. It should also be considered that the solar flux increases manifold within the 200{--}400~nm band during flaring events; 2. The rejection of out-of-band solar flux for wavelengths outside the 200~nm{--}400~nm regime; 3. Reduce the thermal load inside the optical cavity; 4. Allow sufficient light for the darkest features (eg. Sunspot umbrae in Mg lines having $\frac{1}{10}$ brightness of quiet Sun regions) to have a signal-to-noise ratio of at least 100. 

The primary requirement for the attenuation of the thermal filter is defined by the SUIT CCD. The CCD has a full well capacity of 190 ke$^{-}$ ( SUIT ICD document, Issue 3) and a pixel response non-uniformity of 1.5\%. For observing quiet Sun and other darker features, the photoelectron counts are to be maintained between 10 to 110 ke$^{-}$. This provides sufficient dynamic range while leaving sufficient buffer beyond 110 ke$^-$ (60\% full well capacity) to record flaring events without saturating the pixels.

The solar flux varies by a factor of 20 under non-eruptive conditions within 200{--}400~nm. Figure~\ref{spectra} shows the solar spectrum between 0{--}3000~nm, as recorded by the American Society for Testing and Materials \cite{astm_spectrum}. The superimposed red box highlights the SUIT operational wavelength range, further blown up in the inset image shown in the upper right panel. The lower inset shows the solar spectral irradiance between 3000{--}10000~nm.

\begin{figure}
    \centering
    \includegraphics[trim=1cm 3cm 1cm 3cm, width=0.8\linewidth]{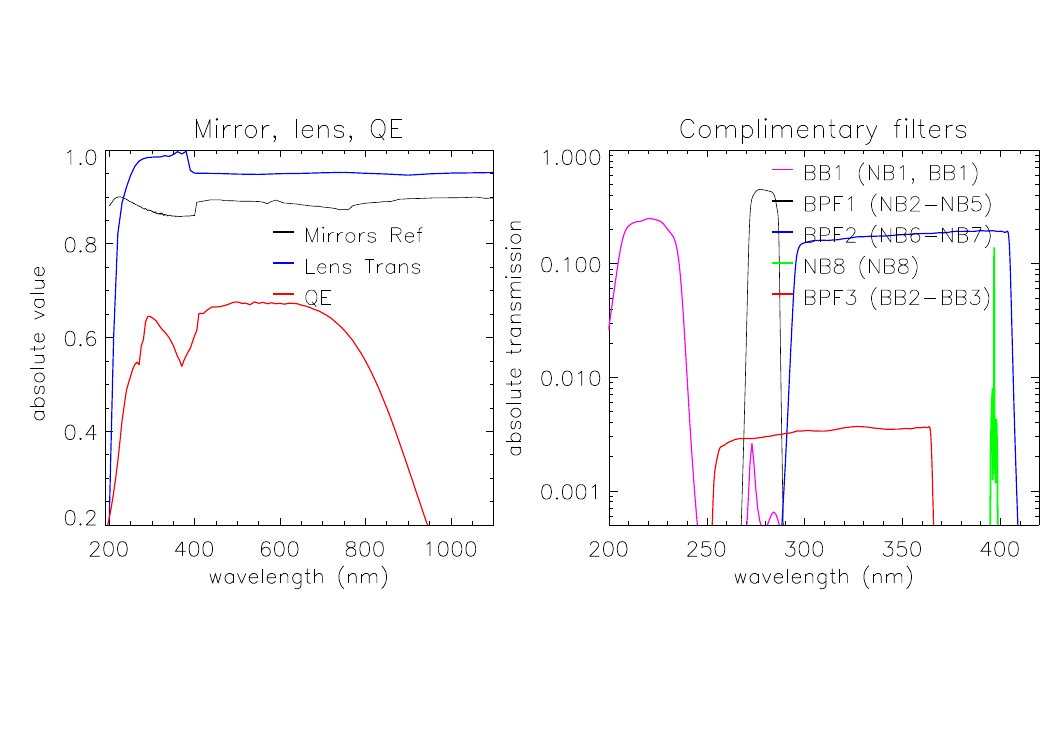}
    \caption{Response curves for all optical components SUIT. The left panel shows the combined reflectance curves for the primary and secondary mirrors (black), transmission curves for the field corrector lens (blue), and the quantum efficiency of the CCD (red). The right panel shows the transmission curves for various complimentary filters of \suit \cite{th_filt_spie}.}
    \label{fig:profiles}
\end{figure}

In addition to that, the FWHM of the science filters vary by a factor of 580 (refer to Table~\ref{scfil}) from the narrowest (narrowband) to the broadest (broadband) filter. This, along with a 20 times increment in solar flux at 400 nm compared to 200 nm, the resultant dynamic range is $\sim$12000. The thermal filter is designed to provide a common attenuation. A set of complementary filters facilitates the remaining cut down for the different science filter bands due to filter-to-filter FWHM variation and change in solar intensity with wavelength. 
In the left panel of Figure~\ref{fig:profiles}, we plot the reflection profile of the mirror(s), the transmission profile of the field corrector lens (FCL), and the quantum efficiency (QE) of the CCD. The transmission profiles of the complementary filters are shown in the right panel of the Figure. The legends indicate the complementary filters along with their coupling science filters. Note that NB01, BB01 and NB08 are coupled with another science filter for optimized performance. 

With these given profiles of the optical components, and the well-defined incoming solar fluxes for each of the science filters, we derive that a minimum rejection factor of $\sim$170 (for non-eruptive events) is required to be achieved by the thermal filter (refer to Table~\ref{cutdown}) within the operational wavelength range. Further additional reduction of fluxes is required by all the other filters. Therefore, an optimized cutdown for all the science filter and complementary filter pairs is achieved with the thermal filter transmission for the exposures between 100{--}1400~ms.

\begin{table}
	\centering
		\caption{The SF identifiers along with the ratio of the photoelectron count generated from the incoming solar flux coming within the PB of each per unit time through the aperture area with respect to the FWC. The third column shows the respective rejection ratio of each SF when the maximum allowed count is 110000. These values are without the thermal filter or the respective CF in the ray path for each SF.}\label{cutdown}
		\vspace{0.1cm}
		\begin{tabular}{@{}llll@{}}
        \hline
			Identification &Eqv. photoelectron flux &Rejection \\
			&/FWC (=190ke$^{-}$s) & ratio\\
            \hline
			NB1 	   &196 & 338\\
			NB2  &185 &320\\
			NB3 &98 &170 \\
			NB4 &129  &223\\
			NB5 &390 6 &674  \\ 
			NB6 &1694 &2927\\
			NB7 &3214 &5551\\
			NB8 &321  &554\\
			BB1 &1011 &1746\\
			BB2 &38460  &66431\\
			BB3 &220280  &380484\\
            \hline
		\end{tabular}
\end{table}

\begin{figure}
	\centering
	\includegraphics[width=1.0\textwidth]{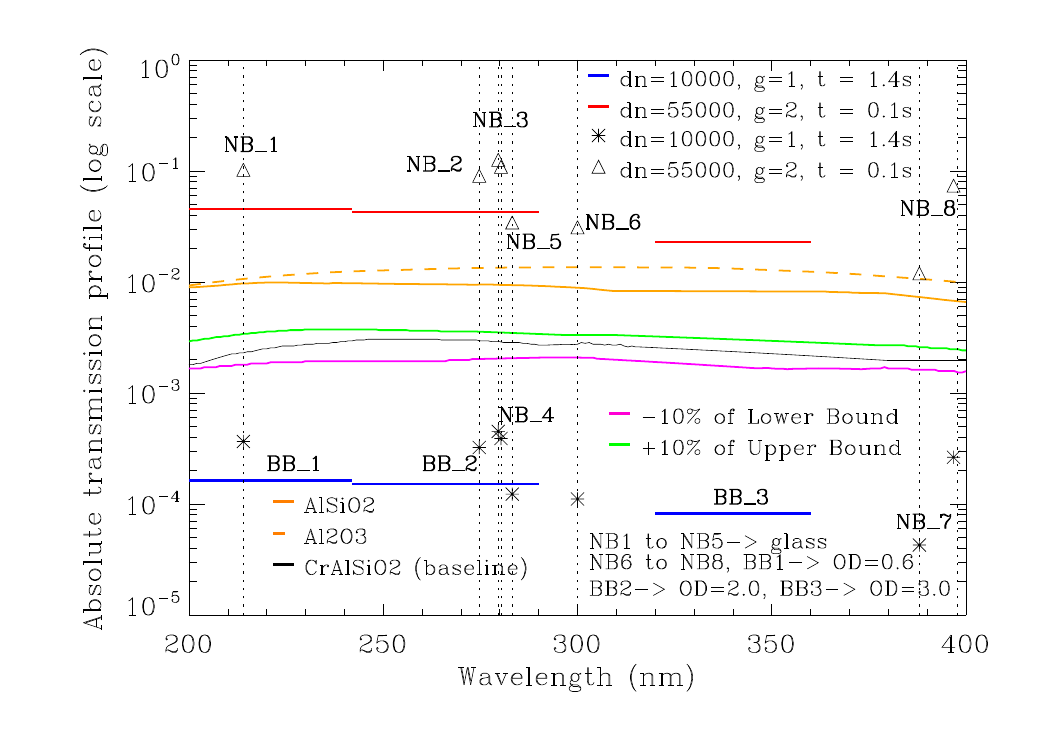}
	\caption{The transmission profile of the thermal filter as a function of wavelength within IB. The red and blue lines show the maximum and minimum transmissions required to achieve the desired number of photoelectrons within a given exposure time (as shown in the legend) for the BBs. The corresponding limits for the NBs are shown as the '$\bigtriangleup$' and '$\ast$' shapes. Overplotted are the transmission profiles manufactured by the vendor for different coating materials and thicknesses in different phases of prototyping \cite{th_filt_spie}.}\label{tf_in}
\end{figure}

Given that the profiles of all optical components are known, except the thermal filter transmission, we limit the maximum and minimum photoelectron count formed on each pixel of the CCD between 110000 at the minimum exposure time of 100~ms and 10000 with the largest allowable exposure time of 1400~ms, respectively. These limits account for the fluxes coming from the brightest as well as darkest features on the solar disc with a contrast ratio of 10:1, with an additional margin of 10\% due to intensity enhancement/ fluctuations. This entire design is aimed for a buffer of $\sim$40\% to reach the Full Well Capacity of 190 ke$^{-}$.
To achieve a throughput within these limits in a given range of exposure time, each of the science filters, along with their complementary filter, sets a range of transmission coefficients of the thermal filter across wavelength. In Figure.~\ref{tf_in}, these are marked by the `$\bigtriangleup$' and `$\ast$', respectively for the NBs. Similarly, for the BBs, the corresponding upper and lower margins of thermal filter transmission are marked in red and blue lines, respectively. In the ambient condition, the thermal filter is expected to have transmission values between the extremes for each science filter-complementary filter combination to accommodate any change in flux as a result of degradation over time under ambient as well as flaring conditions throughout the mission lifetime.

With these design requirements, the final fabrication of the thermal filter has been an extensively iterative process with different choices of substrates and coating materials explained in Section \ref{sec:fabrication}. 

\section{Fabrication and Integration}\label{sec:fabrication}
\begin{figure}
    	\centering
    	\includegraphics[width=0.9\linewidth]{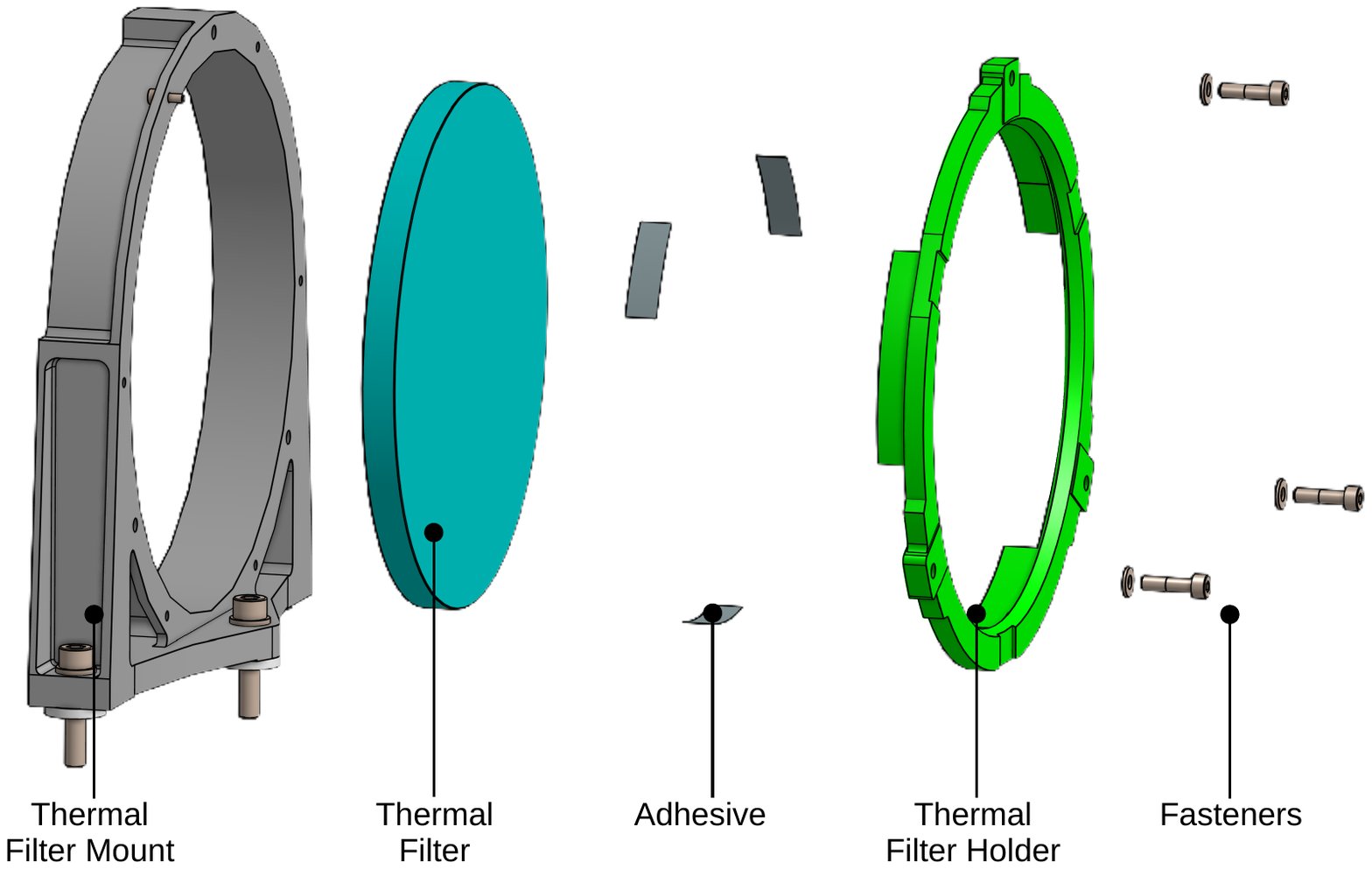}
    	\caption{Exploded view of the TFA, with the Sun-facing side to the left. Individual parts of the assembly are labeled here.}
   	    \label{fig:combo}
    \end{figure}

The Thermal Filter sub-assembly (TFA) consists of three primary parts- the mount, the holder ring, and the thermal filter, as illustrated in Figure~\ref{fig:combo}. Both the holder ring and the mount are made of Ti-6Al-4V alloy. The holder ring has a flexible design to prevent the thermal filter from having stress-induced wavefront errors while meeting simultaneous strength requirements to survive launch and space qualification tests. The thermal filter is glued to the holder ring, which is bolted to the mount. The mount is fastened to the SUIT optical bench with titanium M6 fasteners and required thermal isolation provided by Glass Fiber Reinforced Polymer (GFRP) washers at the fastening points. The details of each component fabrication are explained in the following subsections. 

    \subsection{Thermal Filter}

    \begin{figure}
        \centering
        \includegraphics[width=0.7\linewidth]{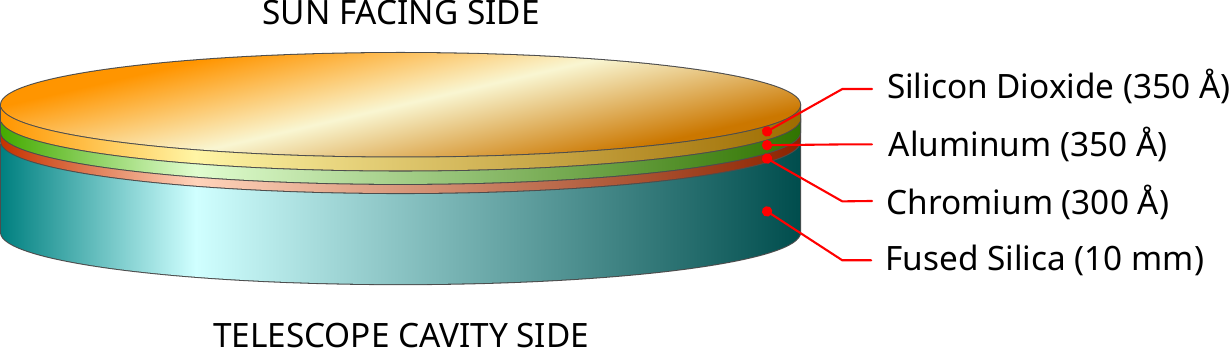}
        \caption{\js{Schematic representation of the thermal filter substrate and the coats applied, along with their respective thicknesses.}}
        \label{fig:tfa_coatings}
    \end{figure}
    
    The thermal filter sits at the front aperture of SUIT, transmitting collimated light from the Sun into the payload optics. Therefore, wavefront errors due to the thermal filter have a significant impact on the instrument's optical quality. Table~\ref{tab:th_filt_spec} gives the specifications for the substrate.
    
    The coating should be of the right materials and thicknesses to achieve the required throughput in near UV, with low out of band transmission. Selecting the substrate and coating for UV optics is very challenging due to the high risk of contamination. In addition, it is necessary for the substrate and the coating to withstand temperatures of $\approx 120^{\circ}C$, sharp thermal gradients, mild thermal shocks, and radiation damages. The manufactured filter should be resistant to oxidation, and corrosion, and the coating should not develop pinholes under normal operation and handling.

    We chose a high-quality fused silica (Corning 7980 1~{\AA}) of thickness 10~mm as the substrate. The substrate \js{has a 300~{\AA} thick coat of chromium (Cr), followed by 350~{\AA} thick layers of aluminum (Al) and SiO$_2$ each.}
    These are deposited by the electron-gun physical vapor deposition (PVD) technique in a Balzers BAK760 coating machine, with an intermediate cleaning procedure to remove contaminants from the surface.
    \js{A schematic diagram of the coats is depicted in Figure \ref{fig:tfa_coatings}.} Further details of the design and coating iterations can be found in \cite{th_filt_spie}. Witness coupons are made, and the coating is qualified for space flight (\S \ref{sec:qualification}).
    The SUIT thermal filter is fabricated after the coupons are seen to meet the required standards discussed earlier in this section. The design and the achieved values for each parameter of the thermal filter are presented in Table \ref{tab:th_filt_spec}.
    
    \begin{table}
    	\centering
    	\begin{tabular}{@{}lllllll@{}}
    		\hline
    		Sl. No. & Parameter & Design Value & Filter \#1 & Filter \#2 & Filter \#3 & Filter \#4 \\
    		\hline
    		1 & Physical Diameter (mm) & 156 (+0.0/-0.1) & 155.95 & 155.95 & 155.96 mm & 155.96 mm\\
    		2 & Clear Aperture (mm) & 152 (+0.5/-0.0) & 152.5 & 152.5 & 152.5 & 152.5\\
    		3 & Thickness (mm) & 10 $\pm$ 0.1 & 10.05 & 10.06 & 10.05 & 9.98\\
    		4 & Testplate fit & \js{$\lambda/4$ PV} at 632 nm & Comply & Comply & Comply & Comply\\
    		5 & Surface Irregularity & \js{$\lambda/4 $ PV} at 632 nm & Comply & Comply & Comply & Comply\\
    		6 & Wedge & $\pm 10"$ & $<10"$ & $<10"$ & $<10"$ & $<10"$\\
    		7 & Surface Quality & 40-20 scratch-dig & Comply & Comply & Comply & Comply\\
    		8 & Surface Roughness & $< 1.5$ nm RMS & Comply & Comply & Comply & Comply\\
    			& before coating  &  &  &  &  & \\
    		9 & Bevel & 0.25 mm $\times$ 45 $^\circ$  & Comply & Comply & Comply & Comply\\
    		\hline
    	\end{tabular}
    	\caption{Inspection Report for SUIT thermal filter.}
    	\label{tab:th_filt_spec}
    \end{table}
    
    \subsection{Filter holder and mount}
\begin{figure}
        \centering
        \includegraphics[width=0.7\linewidth]{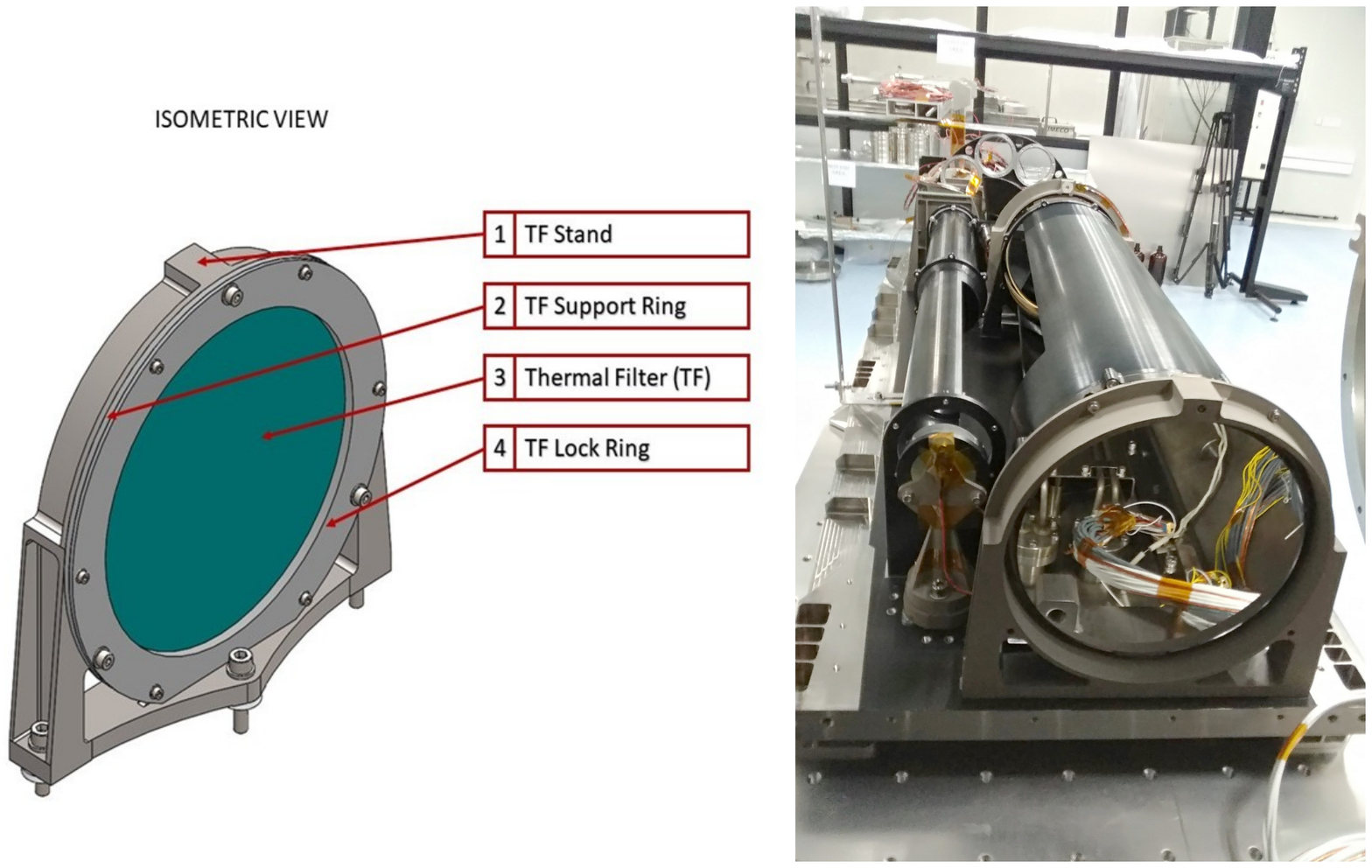}
        \caption{\js{Schematic diagram and photographed of assembled thermal filter assembly \cite{Tripathi2025_SUITMain}.}}
        \label{fig:tfa_photo}
    \end{figure}

    The thermal filter has a diameter of  156 mm and a thickness of 10 mm giving it a diameter-to-thickness ratio of 15.6. A low-thickness glass is chosen to reduce the mass.
    But this comes with the added risk of mechanical and thermoelastic stresses passing onto the glass. A novel design of the holder ring is developed to make the holder flexible enough so that mechanical stresses do not pass onto the filter glass while ensuring it passes spaceflight qualification tests.
    The thermal filter is glued to three flexible blades of the ring, which mechanically isolates the filter from external stresses during vibration and thermal cycling. The thermal filter holder and the mount are made by CNC turning on Ti-6Al-4V blocks.
    The mating surfaces of the holder and mount are fabricated to a flatness accurate to $< 15 \mu m$, to prevent any tilt between the two components.
    It is necessary that there is no tilt between the thermal filter and the filter holder while mounting. Therefore, a special fixture with high flatness is designed and fabricated for gluing the thermal filter to the filter holder. 3M Scotch-Weld Epoxy Adhesive EC-2216 is used for bonding. The adhesive is cured at room temperature for 7 days in a Class 1000 clean-room environment.
    \js{The filter glass is mounted such that the coated side faces the Sun. This is to prevent the unfiltered sunlight from passing through the substrate twice, once during incidence and again during reflection. This could lead to overheating of the thermal filter substrate.}
    After bonding, the thermal filter holder is fastened to the mount with three titanium M4 fasteners at the recommended torque of 275 N-cm, applied uniformly and gradually. The torque levels are derived from the analysis of expected vibration and shock levels from the rocket. The mount is secured to the optical bench with three titanium M6 bolts at the recommended torque of 600 N-cm, with requisite thermal isolation provided by GFRP washers at the mounting points. The reflected wavefront from the sun-facing surface of TF is monitored with an interferometer throughout the mounting process. This is to ensure the thermal filter is not stressed during the gluing and mounting (further details in \S \ref{sec:testing}). Any uneven stresses are relieved by introducing stainless steel shims of appropriate thickness between the mating surfaces. A photograph of the assembled thermal filter assembly is shown in Figure \ref{fig:tfa_photo} \cite{Tripathi2025_SUITMain}.

\begin{figure}
        \centering
        \includegraphics[width=0.5\linewidth]{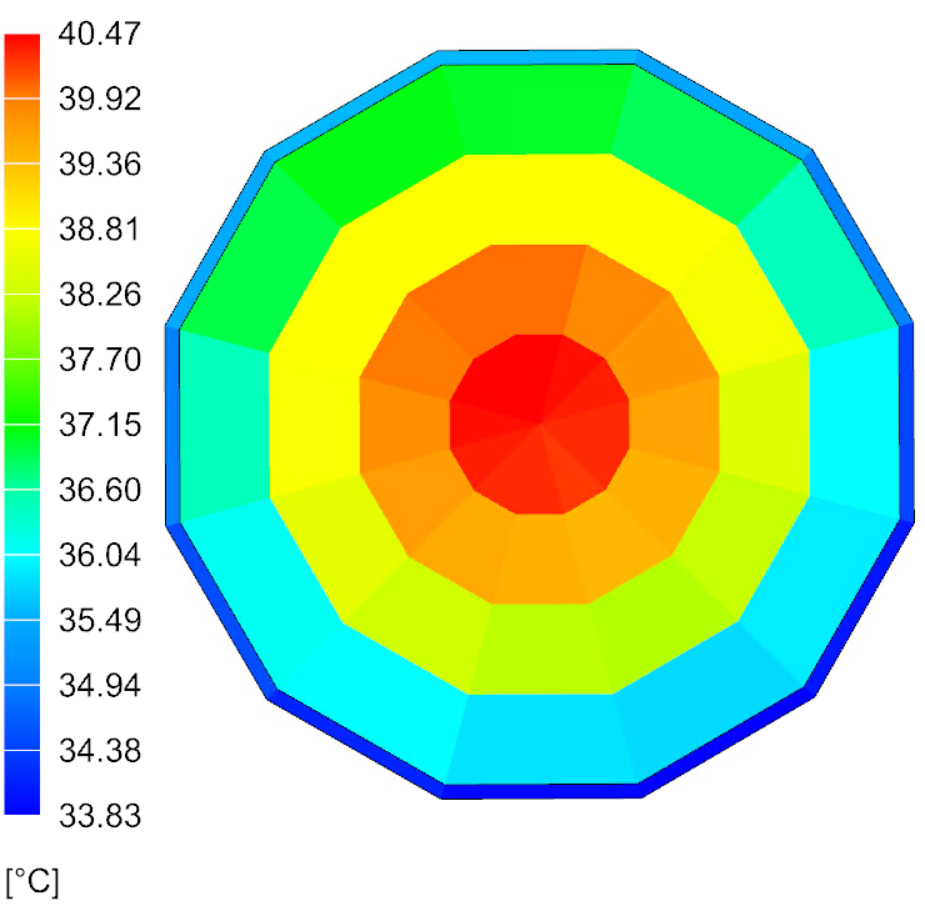}
        \caption{\js{The modeled thermal profile of the sun-facing side of the thermal filter when the assembly is mounted to the SUIT optical bench and is receiving sunlight.}}
        \label{fig:thermal_profile}
    \end{figure}

    From thermal modeling, we find that the thermal filter experiences a temperature of about 43 \degree C in the center and about 30 \degree C at the edges. The gradient along the thickness is comparatively much less. The temperature gradient along the circumference is less than 2 \degree C. \js{Figure \ref{fig:thermal_profile} illustrates the modeled temperature profile of the sun-facing side of the thermal filter when the assembly is mounted to the optical bench.}
    The strong temperature gradient along the radial direction can lead to refractive changes, with the thermal filter working like a thermal lens. The in-flight contribution to optical aberrations by the thermal filter is currently being computed.

    \begin{figure}
        \centering
        \includegraphics[trim=0.cm 1.5cm 0.cm 0.cm, width=0.6\linewidth]{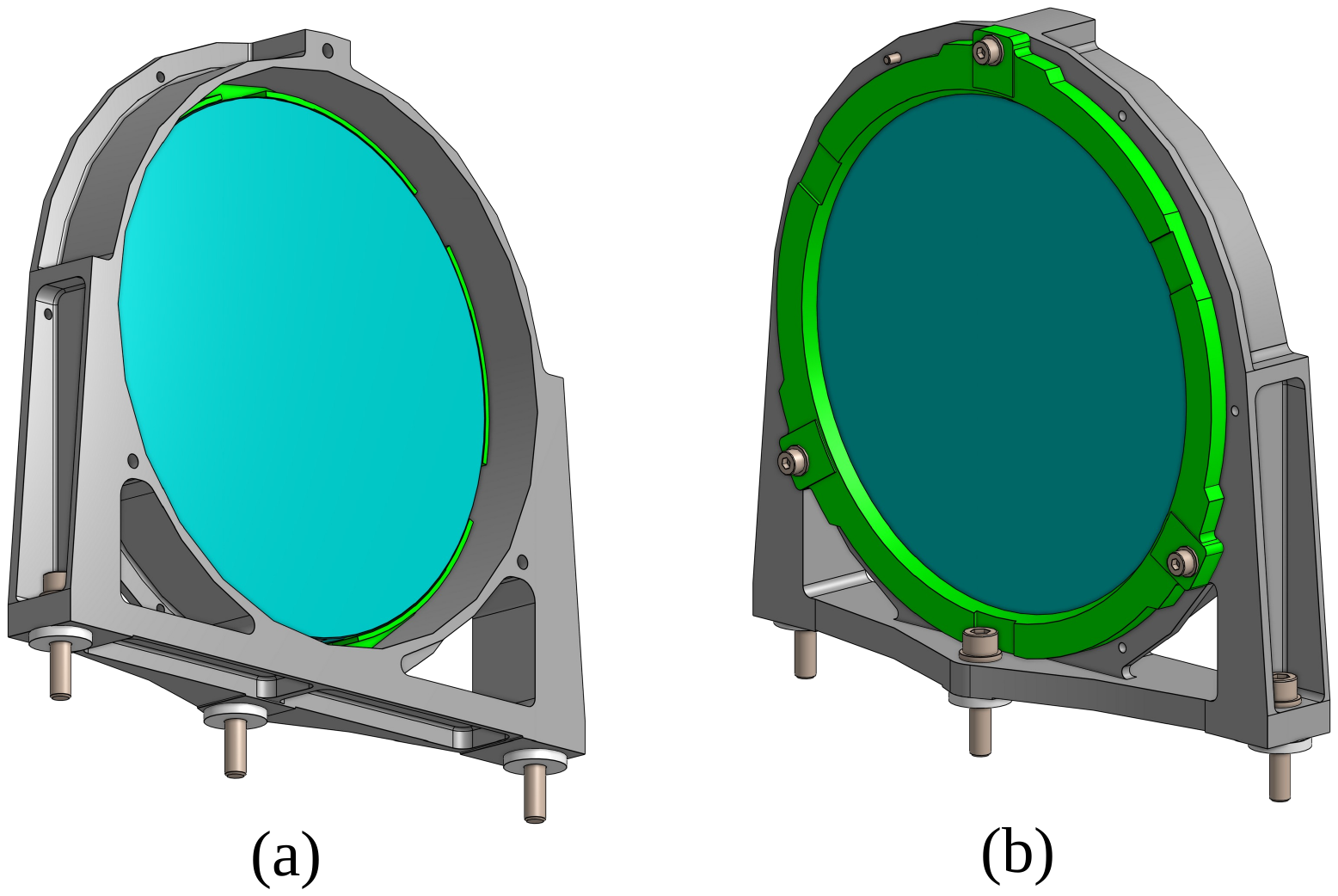} 
        \caption{The figure shows the Thermal Filter Assembly of \suit. (a) Sun-facing side (b)  Payload cavity side.}
        \label{fig:tfa_assembly}
    \end{figure}

\section{Space Qualification} \label{sec:qualification}
\begin{figure}
    \centering
    \includegraphics[width=1\textwidth]{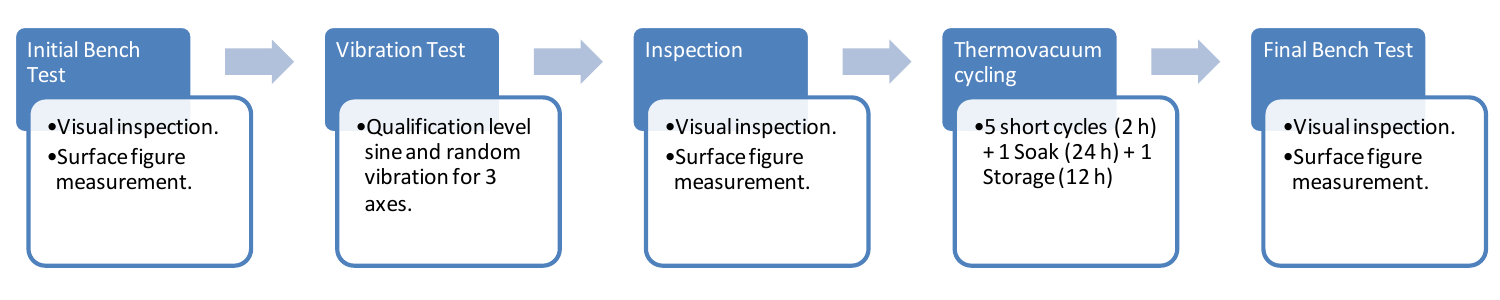}
    \caption{Space qualification flow chart}\label{tfa_qualification}
\end{figure}
    \subsection{Coating qualification}
\begin{figure}
        \centering
        \includegraphics[width=1.0\textwidth]{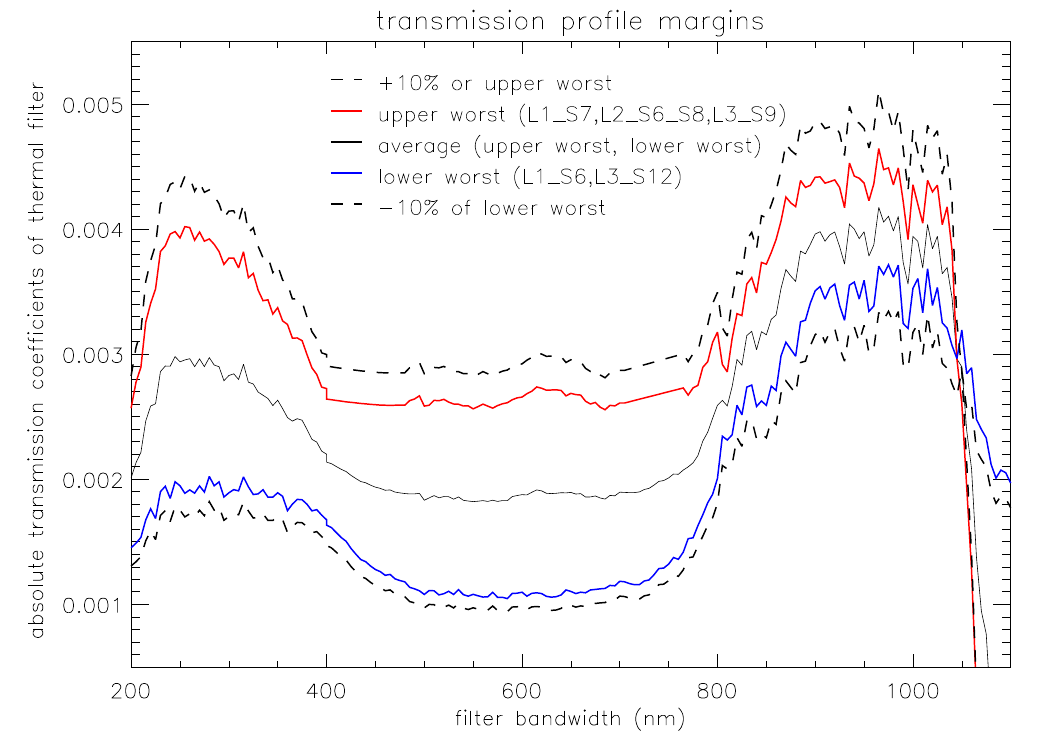}
        \caption{Transmission profiles of the TF as a composite of measurements performed for few of the sample coupons (referred to in the legend). The red and blue curves show the upper and lower values of the transmissions in different wavelength ranges for these samples. The intermediate black curve is an average of the red and blue curves. The dashed black lines indicate an additional $\pm$10\% margin on the thresholds. These dashed curves serve as the qualification thresholds. `L' stands for Lot and `S' for Sample \cite{th_filt_spie}. }\label{uplow}
    \end{figure}

    As a test of variations in transmission and/or reflection properties due to imperfect coating procedures, several small coupons, each of diameter 25~mm, are coated, and measurements are performed at the Laboratory for Electro-Optics Systems of the Indian Space Research Organization. The spectroscopic test is performed within 200{--}1100~nm using a UV/VIS Perkin Elmer Spectrophotometer. These coupons have slightly different transmission values, and a composite of these values at different wavelengths gives us an estimate of the lowest and highest transmission at any given wavelength. These are shown as the red (maximum) and blue (minimum) lines in Figure~\ref{uplow} and serve as the thresholds for qualification tests of the thermal filter. An average of these extreme transmission profiles is plotted as the black solid curve in the same plot. This is noted to be of similar values to that in the `baseline' design profile. Another $\pm$10\% margin on these extreme transmission profiles (two dashed lines in Figure~\ref{uplow}) also doesn't lead to any violation of the throughput and signal-to-noise limits set as the design parameters. This proves the robustness of the performance of the thermal filter. 
    
\begin{figure}
        \centering
        \includegraphics[width=1.0\textwidth]{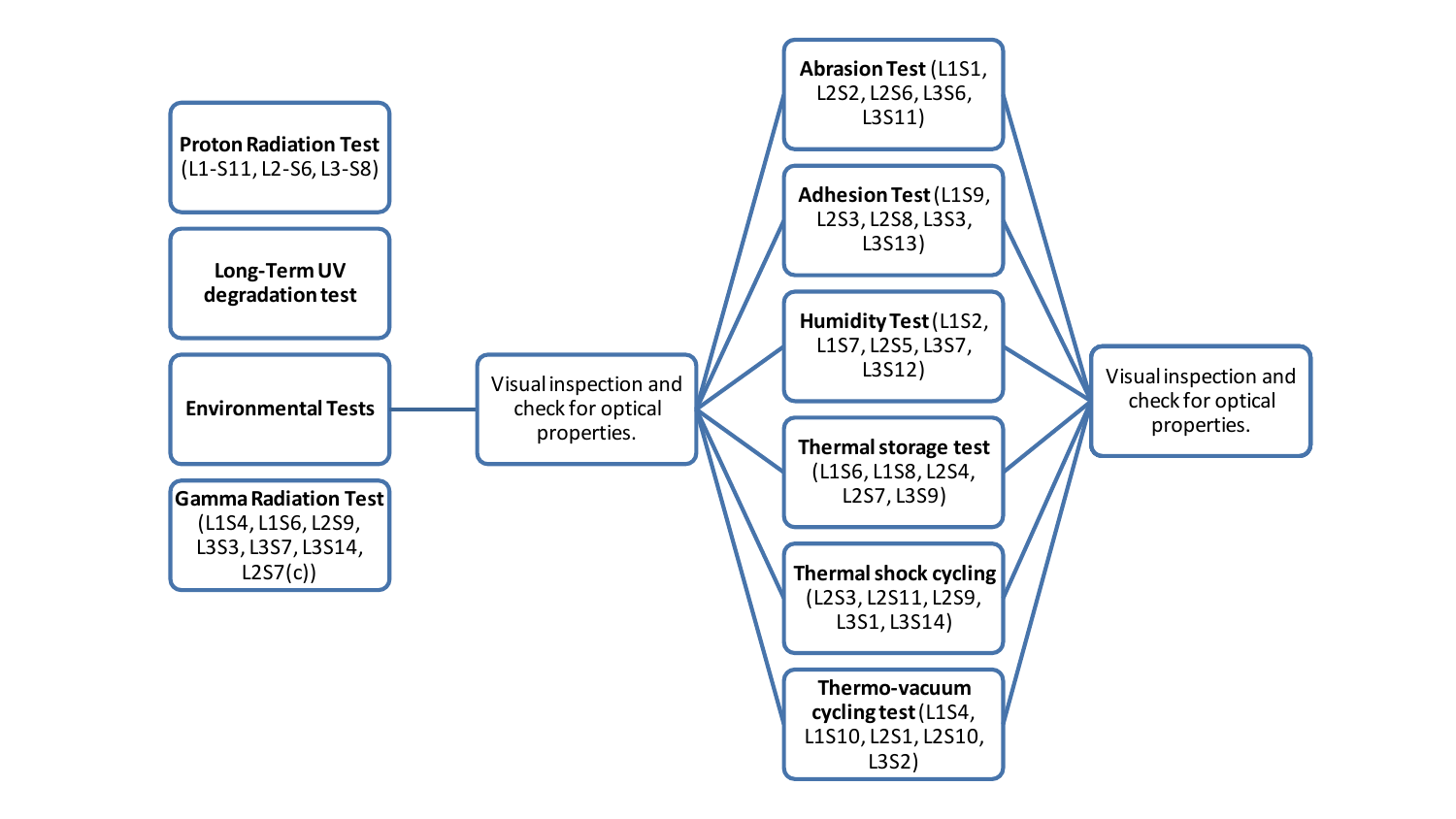}
        \caption{Flowchart showing the details of the qualification tests, including the coupons identifiers used for the purpose. Here, `L' and `S' denote Lot and Sample number, respectively. L2-S7(c) indicates the control coupon, wherever applicable.}\label{flow}
    \end{figure}

    Since the SUIT TF is located at the entrance aperture, the highly reflecting coated surface is exposed to the space environment throughout the mission lifetime. Therefore, it is imperative that it is subjected to rigorous qualification tests. In Fig~\ref{flow}, we note the details of the tests performed and the coupons that have been used for various tests. Here, `L' and `S' denote Lot and Sample number, respectively, which essentially indicate different batches of coating activity. After each test, a visual inspection is done with the naked eye under 100~W illumination and a Nikon Profile Projector. This is followed by a spectral characterization between 200{--}800~nm with a step size of 5~nm using UV/Vis Perkin Elmer Spectrophotometer.
    The space qualification tests for the TF include environmental as well as gamma and proton radiation tests. In the first stage, the environmental tests are performed, comprising thermal storage (TS), thermo-vacuum (TVac), humidity, thermal shock cycling (TSC), adhesion, and abrasion tests. 
    
    Subsequently, the gamma radiation test is performed with three different amounts of radiation - 1000, 1500, and 3250 Krad, at the radiation facility at the U.R. Rao Satellite Centre of ISRO. This is based on a model of input radiation dosage rate of 168 rad/s, incident on the TF top surface of thickness 400~{\AA} at the L1 position of the Sun-Earth system, performed with the SPENVIS software. In Table~\ref{dose}, we indicate these radiation values, the corresponding coupons, and the exposure times required for each of the radiation values. 
    \begin{table}
    \centering
            \caption{Table for the time required to expose the coupons to radiations of 1000, 1500 and 3250 Krad with the input rate being 168 rad/s. Here `L' and `S' denote Lot and Sample number respectively.}\label{dose}
            \begin{tabular}{@{}lll@{}}
                \hline
                Radiation (kRad) & TF coupon & Exposure (mins) \\ 
                \hline
                1000 &	L1-S6, L3-S14 &100 \\
                1500 &L3-S2, L2-S9 &150\\
                3250 &L1-S4, L3-S7 &325 \\
                \hline
            \end{tabular}
    \end{table}    
    
    \begin{figure}
        \centering
        \includegraphics[width=1\textwidth]{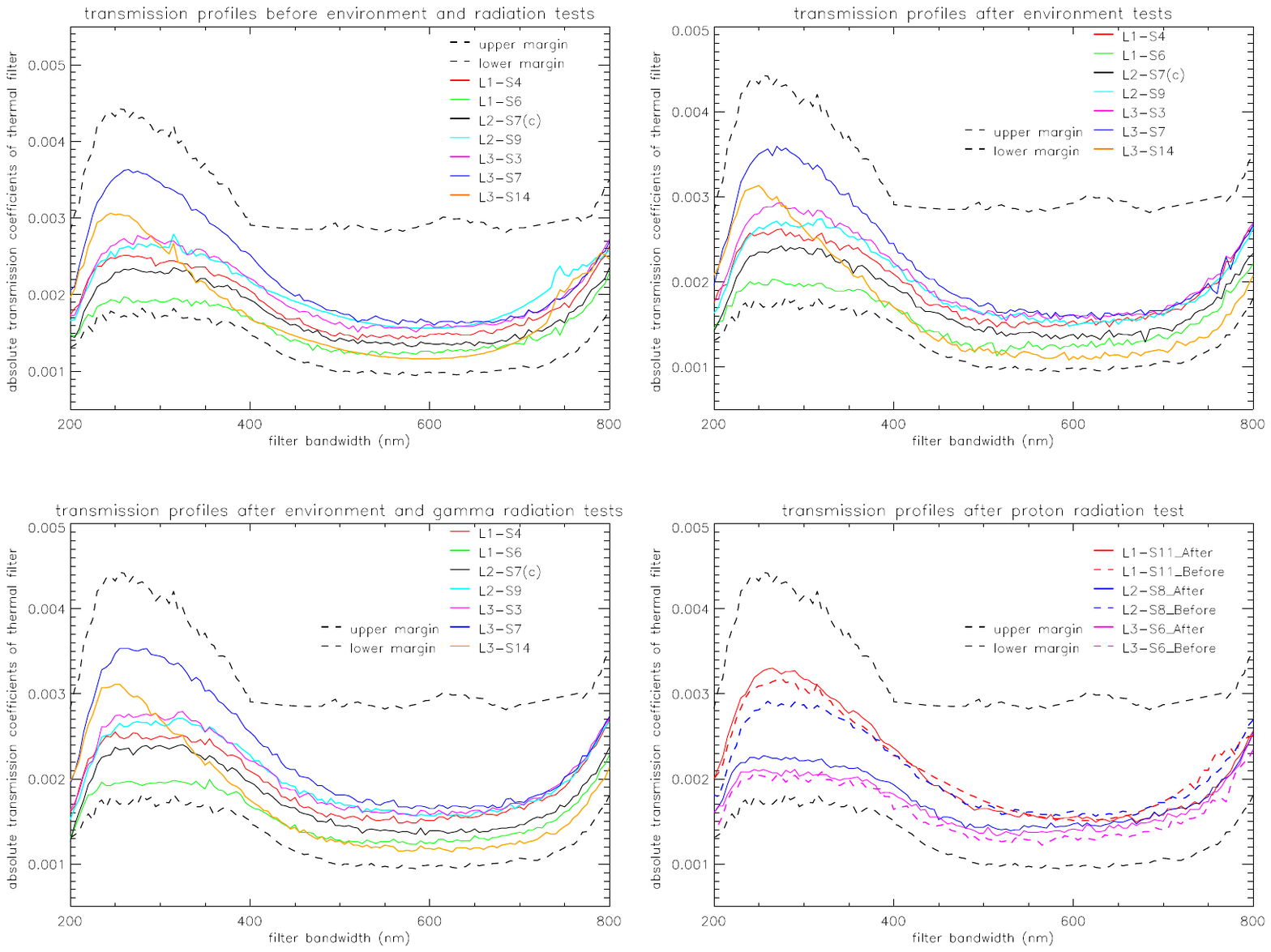}
        \caption{Top Left panel: Transmission profiles of seven coupons before any of the environmental or radiation tests are conducted. Top Right panel: Transmission profiles of seven coupons (colored lines) after all of the environmental tests are conducted. Bottom left panel: Transmission profiles of seven coupons after all the environmental and gamma radiation tests are conducted. In all three panels, these respective profiles are shown in colored lines, whereas the upper and lower allowed margins are indicated by the dashed black lines. L2-S7 is the control coupon. Bottom right panel: Transmission profiles of three thermal filter coupons before (red, green and magenta dashed curves) and after (red, green and magenta bold curves) the proton radiation test. It is to be noted that the L1-S11 and L2-S8 samples have been exposed to 100\% fluence whereas L3-S6 is exposed to only 50\% fluence. The permissible range of the transmission profiles are also indicated by the dashed black lines \cite{th_filt_spie}.}\label{fig:coupons}
    \end{figure}
    
    The two panels on the top and bottom left panel in Fig.~\ref{fig:coupons} show the transmission profiles of these coupons post environmental and, thereafter, gamma radiation tests. The top left panel shows the same for six coupons before any of the tests are performed. The top right panel shows the same after environmental tests, followed by the bottom left panel representing post-environmental-gamma-radiation tests. L2-S7 is the control coupon for these tests. The dashed black lines indicate the acceptable upper limits of transmission after all the aforesaid tests are done. We note that the transmission curves before and after these qualification tests are well within our acceptance range.  
    
    In the following step, we perform the proton radiation test. As in the gamma radiation test, we use the SPENVIS software to compute that a fluence of 6.4 $\times$ 10$^{12}$ protons/cm$^{2}$ of 21.5 MeV protons could be incident on the TF in the lifetime of the telescope. At the BARC-TIFR Pelletron facility at TIFR, Mumbai, we perform this test on three coupons (L1-S11, L2-S8, L3-S6). For further details, refer to Table~\ref{prot}. Here, we would like to emphasize that L3-S6 is restricted to $\sim$50\% exposure only (3.21 $\times$ 10$^{12}$ protons/cm$^{2}$) whereas, L1-S11 and L2-S8 are exposed to nearly 100\% fluence (6.43 and 6.41 $\times$ 10$^{12}$ protons/cm$^{2}$).
    
    \begin{table}[h]
        \centering
            \caption{Specifications for the proton radiation test conducted at the BARC-TIFR Pelletron facility. Here `L' and `S' denote Lot and Sample number respectively.}\label{prot}
            \vspace{0.1cm}
            \begin{tabular}{| c | c |} \hline
                Radiation &Thermal filter \\ 
                level (\%) &coupons used \\ \hline
                100 &	L1-S11  \\ \hline
                100 & L2-S8 \\ \hline
                50 &L3-S6 \\ \hline
            \end{tabular}
    \end{table}    
    
    In the bottom right panel of Fig.~\ref{fig:coupons}, we show the initial as well as final transmissions of the three coupons by the dashed and bold curves, respectively. The permissible limits of the TF transmission values are the same as considered for the environmental and gamma radiation tests (refer to Figure~\ref{fig:coupons}). From Figure~\ref{fig:coupons}, we note that two of the coupons (L1-S11 and L3-S6) show a maximum change of about 7\% in transmission between 200{--}400~nm range whereas it reduces to less than 5\% in 400{--}800~nm band. When noted with care, we see that the transmission profile shows a slight increase in the post-radiation phase for these two samples. This could well be due to measurement errors at the LEOS facility, but within acceptable limits. However, the transmission decreases by 25\% for L2-S8 after proton irradiation between 250{--}300~nm domain. The decrease is pertinent at wavelengths outside this range ($<$10\%). It is noted that the post-proton radiation test transmission curves of these three thermal filter coupons lie fairly within the extreme profiles that have been generated after the environmental and gamma radiation tests.
    
    \subsection{Assembly Dynamical tests} \label{subsec:dynamical_test}
    
    Assembly-level dynamical tests were performed to qualify the TFA for launch environments. The input vibration levels were provided by the flight dynamics team. The dynamic qualification is carried out by performing Finite Element Method (FEM) simulations and vibration testing. The locations on the assembly with high dynamic acceleration and critical to optical performance were noted during the simulation. Some of these regions were used for accelerometer sensor placement during vibration testing. The vibration tests are done on all three axes. Two vibrations each were performed in-plane and out-of-plane, the levels for which are listed in Table~ \ref{tab:inplane} and \ref{tab:Outofplane}.
    The vibration testing is carried out for both sinusoidal and random vibration loads. A low-level resonance sinusoidal search is carried out both before and after the vibration tests to validate the performance and workmanship of the assembly. For each axis, low-level, medium-level, and qualification-level vibrations are applied. Similar steps are repeated in the other two axes as well. 
    
    The first mode for vibration along the optical axis is seen at 272 Hz. It is also seen that there is no change in the peak response observed before and after the tests. This demonstrates that the thermal filter assembly is qualified for a flight dynamic environment. In addition, optical tests are also performed before and after dynamical tests, to ensure the optical properties of the filter are intact. 
    A low-level random (LLR) vibration test is performed on the critical axis, and the responses from all accelerometers are studied to confirm the simulated estimations. This is followed by qualification level vibrations in sinusoidal and random to the specified levels. LLR is repeated, and the response curves are compared with the pre-test LLR response curves to ensure no structural changes occurred during the vibrations.  
    All through these tests, an extreme contamination control protocol is followed.
    
    \begin{table}
        \centering
        \begin{tabular}{||c|c||c|c||}
            \hline
            \hline
            \multicolumn{2}{|c|}{In Plane - Sinusoidal} & \multicolumn{2}{|c|}{In Plane - Random} \\
            \hline
            Frequency (Hz) & Amplitude (g) & Frequency (Hz) & PSD ($g^2/Hz$)\\
            \hline
            5-20 & 9.3 mm (O-P) & 20-100 & +3 dB/oct\\
            20-50 & 20 & 100-700 & 0.15\\
            50-70 & 8 & 700-2000 & -3 dB/oct\\
            70-100 & 2 oct/min & $g_{RMS}$ & 14\\
            Sweep Rate & 1 & Duration & 2 min\\
            Num of sweeps & 1 &  & \\         
            \hline
            \hline
        \end{tabular}
        \caption{Test levels for `In Plane' vibration test of \suit TFA.}
        \label{tab:inplane}
    \end{table}

    \begin{table}
        \centering
        \begin{tabular}{||c|c||c|c||}
            \hline
            \hline
            \multicolumn{2}{|c|}{Out of Plane - Sinusoidal} & \multicolumn{2}{|c|}{Out of Plane - Random} \\
            \hline
            Frequency (Hz) & Amplitude (g) & Frequency (Hz) & PSD ($g^2/Hz$)\\
            \hline
            5-20 & 12.4 & 20-100 & +3 dB/oct\\
            20-50 & 25 & 100-700 & 0.33\\
            50-70 & 15 & 700-2000 & -6dB/oct\\
            70-100 & 8 & $g_{RMS}$ & 19\\
            Sweep Rate & 2 oct/min & Duration & 2 min\\
            Num of sweeps & 1 &  & \\    
            \hline
            \hline
        \end{tabular}
        \caption{Test levels for `Out of Plane' vibration test of \suit TFA.}
        \label{tab:Outofplane}
    \end{table}
    
    \begin{table}
        \centering
        \begin{tabular}{||c|c||}
            \hline
            Axis & Amplified frequency(ies) (Hz) \\
            \hline
            X & 272.5\\
            Y & 461.8, 729.6\\
            Z & 1193.9\\
            \hline
        \end{tabular}
        \caption{The dynamic amplification frequencies of the thermal filter vibration test. The three axes are the responses from the three orthogonally placed accelerometers. The first resonance mode is well above the critical limit of 100 Hz.}
        \label{tab:tfa_vib}
    \end{table}
    
    \subsection{Thermal Filter Assembly Thermovacuum tests} \label{subsec:tvac_test}
\begin{figure}
        \centering
        \includegraphics[width=0.8\textwidth]{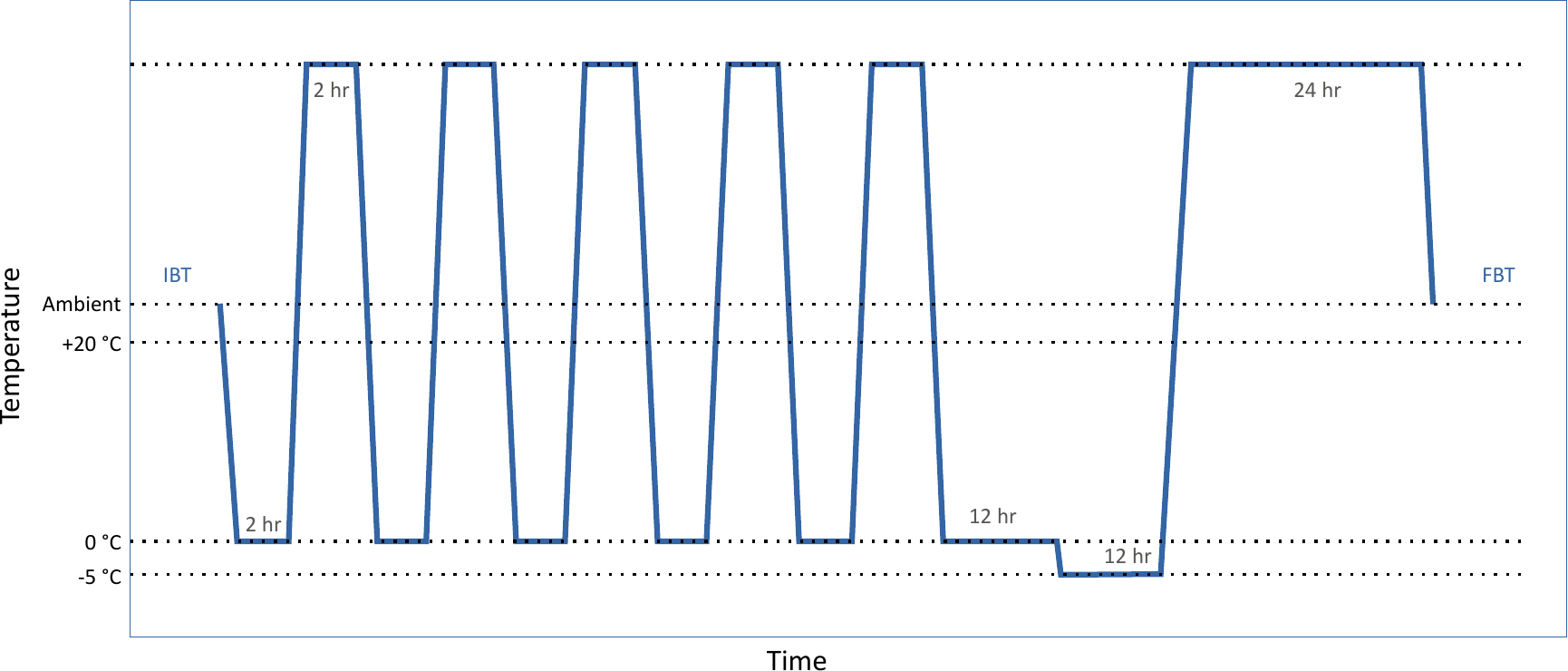}
        \caption{Temperature level specifications of thermovacuum cycling tests.}\label{fig:tvacprofile}
    \end{figure}

    After successfully completing dynamical testing, the assembly is loaded into a thermo-vacuum chamber for thermal storage and cycling tests. The thermal profile is derived based on expected thermal environments during the lifetime of the assembly in space environment during operations and in storage. Sufficient margins are added to the levels while deriving the test profile, as depicted in Figure~\ref{fig:tvacprofile}. Optical tests are performed pre and post-thermovacuum cycling tests to confirm the performance.
    
    The contamination control protocol demanded stringent cleanliness conditions for the vacuum chamber used for these tests. To achieve this, the chamber alone was baked continuously for 72 hours, and purged multiple times with high-purity gaseous nitrogen. The outgassing levels of the chamber were verified using a TQCM (Thermoelectric Quartz Crystal Microbalance) system. The chamber cleaning is continued till the specified limit of $10 ng/Hr/cm^2$ of mass deposition is achieved. The TQCM monitoring is continued during the tests as well to make sure there is no outgassing from the assembly, indicating contaminant deposition on the assembly.
    
    \subsection{Accelerated coating life test}
    An accelerated life test is performed to check how the thermal filter coating degrades with time and solar exposure. Light from a sun-simulating xenon lamp is focused with a UV-fused silica lens. The intensity of the light after focusing is measured with a NIST traceable photodiode. The net flux is measured to be three times that of the total solar radiation per unit area that the filter would be exposed to in the space environment. Two filter coupons are mounted inside a 0.3 m vacuum chamber. One of the coupons sits at the focus of the light beam from the xenon lamp, while the other is kept away from the light beam. It is ensured that the light from the lamp is incident normally on the filter coupon. The vacuum chamber was maintained at a pressure of $\sim 10^{-3}$ bar, and the filter was exposed to the light from the lamp. Before the test is started, a spectroscopic transmission test is done for both coupons.

    The ratio of the transmission spectra through the exposed coupon and the unexposed coupon is taken to see the relative variation in transmission for the exposed coupon.
    We notice that the transmission factor changes from from 1.0 to 1.14 at 400 nm after 60 hours exposure (effectively 180 hours of solar exposure). However, the ratio increases to 1.17 after 163 hours of exposure (effectively 489 hours of solar exposure). So the degradation slowed down with time, which would eventually settle to a negligible amount in space.
    
\section{Optical testing and characterization}\label{sec:testing}
\begin{figure}
    \centering
    \includegraphics[width=0.7\textwidth]{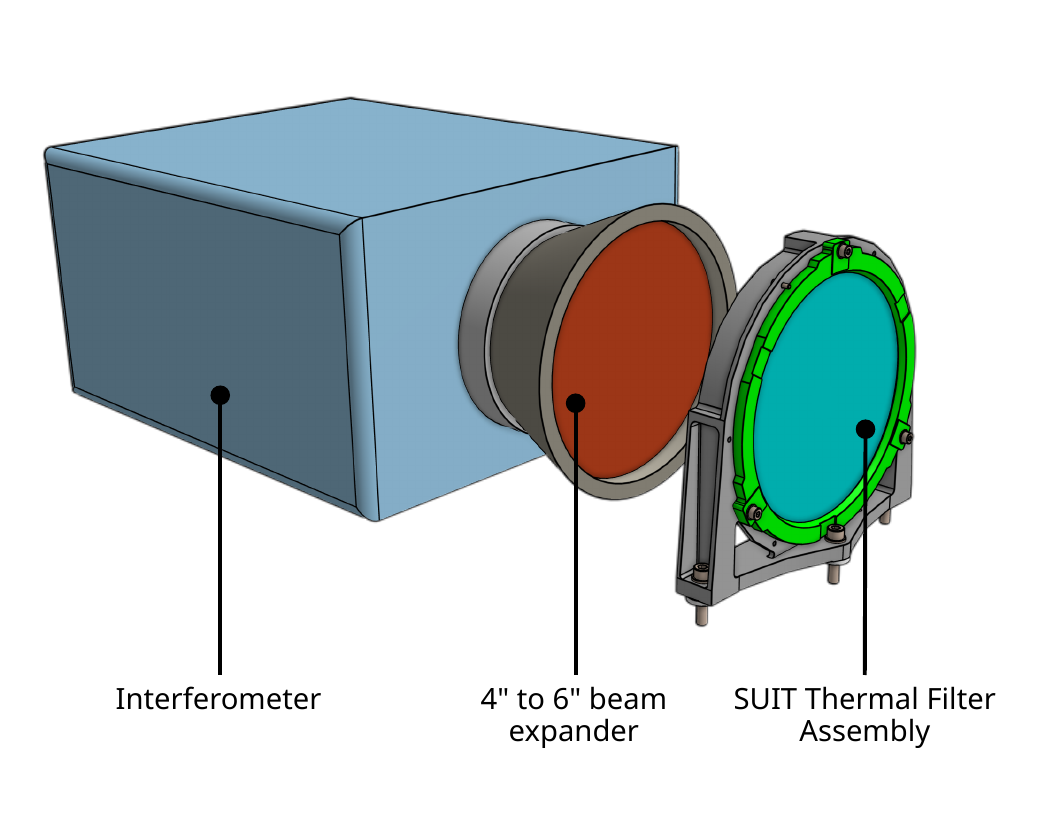}
    \caption{Schematic diagram of the wavefront measurement setup for SUIT Thermal Filter}\label{metrology_setup}
\end{figure}

Optical testing of the Thermal Filter is carried out in two phases- pre and post-coating. The transmission wavefront of the glass blank complies with the requirement of \js{PV $\lambda/4~at~632 nm$.} The wedge angle of the blank is lesser than $\pm 10"$ with a surface roughness of less than $1.5 nm$ RMS. The blank is beveled at 45 degrees by 0.25 mm.

Post-coating, it is not possible to measure the thermal filter's transmission wavefront because of $\sim 0.01 \%$ transmittance at wavelengths longer than 600 nm. The transmitted light fraction is insufficient to perform interferometry with reliable SNR and contrast. Instead, interferometry is performed with the reflected wavefront from the coated surface with a 4-inch Zygo Verifire Fizeau Interferometer equipped with a 4-inch to 6-inch beam expander as shown in Figure \ref{metrology_setup}. 

\begin{figure}
    \centering
    \includegraphics[width=0.8\textwidth]{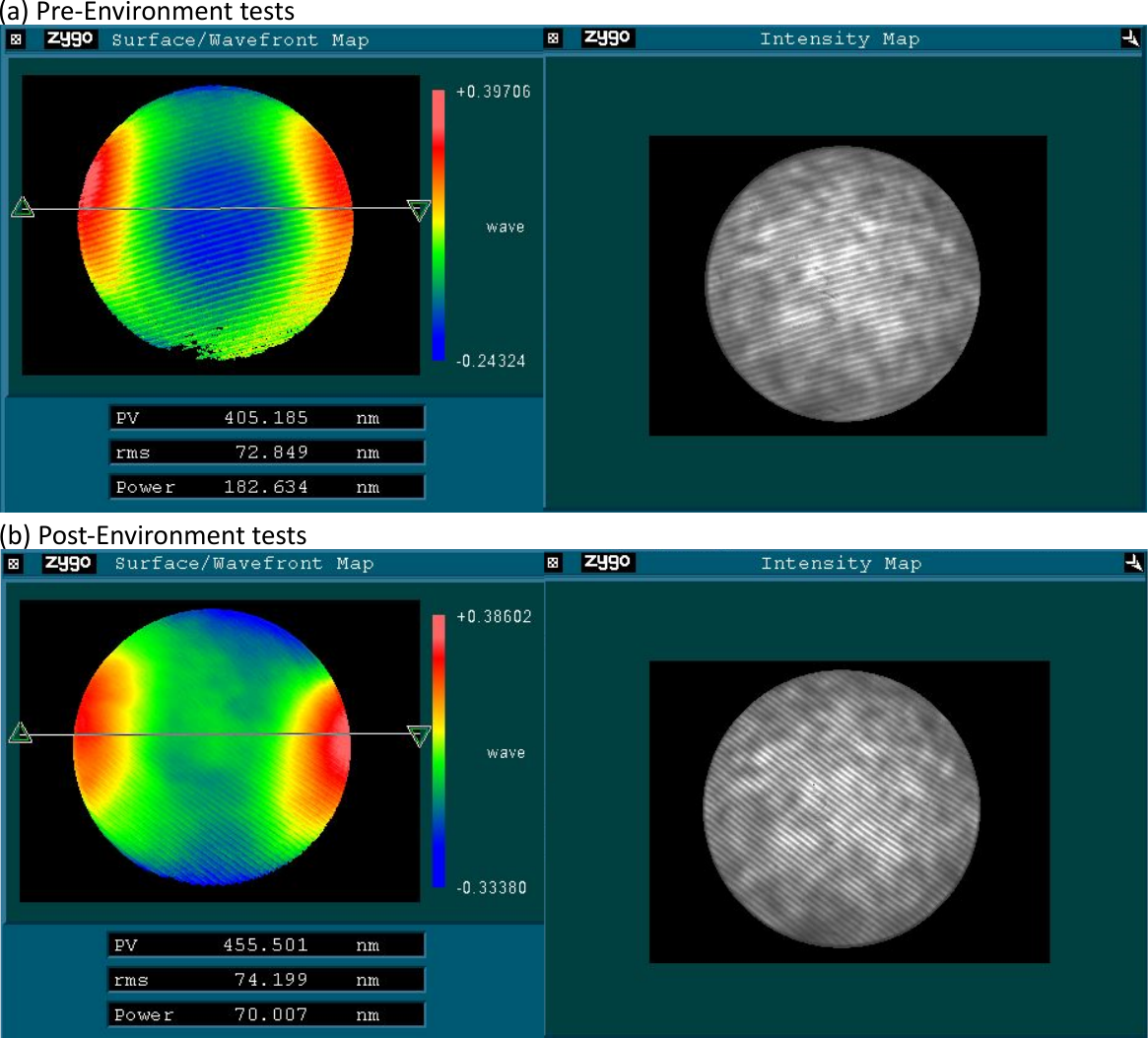}
    \caption{Surface phase profile of the Thermal Filter assembly measured using an interferometer pre (top) and post (bottom) environmental tests. The RMS wavefront error did not change significantly.     
    The power observed before environment tests was redistributed and settled to a lower value of $70nm\pm 2 ~nm$ after the environment tests.} \label{fig:tfa_wavefront}
\end{figure}

The coated thermal filter is mounted on the frame with the fasteners uniformly tightened at 100 N-cm, which is less than half the recommended torque. Interferograms are recorded as the bolts are gradually tightened. Uneven clamping force introduces stress-induced curvature on the glass. This leads to stress-induced wavefront aberrations in the reflected wavefront. Under such a scenario, stainless steel shims are added between the frame and the ring to evenly distribute the stress on the thermal filter. All the fasteners are finally torqued at 274 N-cm. 
After integration, the TFA went through all the environmental tests described in \S \ref{subsec:dynamical_test} and \S \ref{subsec:tvac_test}. The optical performance of the reflected surface is monitored at each stage of the environmental test as a verification check for qualifying the process. The thermal filter is considered acceptable when the RMS (root mean square) error of the reflected wavefront is less than $\lambda/8$ at 632.8 nm, and it does not change drastically after environmental tests. Figure \ref{fig:tfa_wavefront} shows the surface phase profiles of the reflected surface of the TFA before and after the environmental tests.

\section{Conclusions and Discussions}

The Solar Ultraviolet Imaging Telescope (SUIT) on-board the Aditya-L1 satellite is an imaging telescope operating in the Near-UV (200{--}400~nm) domain of the solar spectrum, taking high cadence and spatially resolved images from the lower photosphere to the upper chromosphere. 11 science filters, coupled with appropriate complementary filters are used for imaging. 8 science filters address the questions of coupling of mass and energy as well as different eruptive events on the Sun. The remaining three filters aim to understand the role of NUV emission from different features of the Sun in space-terrestrial weather. 

Considering the incoming solar flux, we have to ensure that the CCD is in the linearity regime for each SF with further suppression at wavelengths beyond the operational wavelength range of SUIT. Moreover, we aim to constrain the photoelectron flux for sufficiently high SNR, even those for the darkest features on the Sun with a contrast ratio of 10:1 with the bright features, under ambient conditions. In addition, the heat load across the telescope needs to be maintained within 20$\pm$1 \degree C for optimized imaging performance. To achieve these design objectives, we design and fabricate the highly reflective Thermal filter placed at the entrance of the telescope.

The thermal filter substrate material is chosen to be fused silica (Corning 7980 1~{\AA}) with double layers of Cr (300 and 110~{\AA} thick respectively) and then Al and SiO$_{2}$ (350~{\AA} thick). The coating deposition is done in a highly controlled environment using the electron-gun PVD technique.

The TF has an average transmission between 0.1{--}0.45\% between 200{--}400~nm and persists at about 0.2\% in the Visible band (as measured with a Perkin Elmer Lambda 9 spectrophotometer). It is important to mention that though the average transmission coefficients increase in the IR range, the solar flux itself is low in this domain, thereby maintaining the thermal performance restrictions. In the shorter wavelengths, the average transmission is $<$0.2\%. Even with a $\pm$10\% variation in the transmission values, the performance of the telescope is maintained well within limits. Therefore, it can be concluded that the thermal filter significantly reduces the flux entering the system.

The heating up of the subsystems is also prevented by ensuring highly reflective coating layers, robust enough to prevent pinhole formation at the fabrication stage. The reflection is $\sim$45{--}70\% upto 400~nm and $>$80\% thereafter. This results in a total of about $\sim$3~W heating ahead of the primary mirror for incoming solar radiation. 

The TF is placed at the entrance, thereby enduring harsh space environment and radiation conditions while in orbit for the mission lifetime. It is also robust enough to sustain the manufacturing and handling processes as well as the corrosion/oxidation of the coating layers. The coating material needs to be durable enough to withstand the cleaning processes, which are important for contamination-free assembly / high throughput in the UV region.

To verify the performance degradation, several qualification tests, including environmental tests, gamma and proton radiation tests at different fluence values are performed on several samples of the TF coupons. It is confirmed that the transmission profiles of the coupons do not change much after these qualification tests and still comply with the desired design constraints, in terms of transmission. The SUIT is operational onboard Aditya-L1 for more than an year and is working perfectly well. The thermal filter has undergone a real test of rocket launch, cruise to L1 crossing the Van Allen radiation belts, and \js{has completed} around 14 months in halo orbit around L1. The images taken using SUIT, as shown in \cite{Roy_2025}, show acceptable image quality.\js{However, the angular resolution of the instrument falls short of the design specifications as elucidated in \cite{sarkar_test_2025}. The thermal filter is the first and only optical element directly exposed to unfiltered sunlight. We are determining if deviations due to thermally induced stresses on the thermal filter can account for the delivered imaging performance shortfall.}

The significance of an entrance aperture filter with a high reflection coefficient and requisite transmission within 200{--}400~nm lies in the fact that it maintains the photometric performance in terms of throughput and SNR for several filters across the range with variable FWHM and prevention of leaks beyond the operational wavelength range, particularly, with the large flux in the VIS-range immediately following the NUV domain of the electromagnetic spectrum. To the best of our knowledge, this is the first attempt to make such an entrance aperture filter for imaging in the NUV domain (unlike SuFI and IRIS), and it can be used in the future for solar telescopes aiming at observations in similar wavelength domains.

\subsection* {Code, Data, and Materials Availability} 
The data presented in various plots across the paper can be made available upon reasonable request to the corresponding author.

\subsection*{Acknowledgments}
\js{We thank the reviewers for the constructive comments and suggestions.} {\suit} is built by a consortium led by the Inter-University Centre for Astronomy and Astrophysics (IUCAA), Pune, and supported by ISRO as part of the Aditya-L1 mission. The consortium consists of SAG/URSC, MAHE, CESSI-IISER Kolkata (MoE), IIA, MPS, USO/PRL, and Tezpur University. Aditya-L1 is an observatory class mission that is funded and operated by the Indian Space Research Organization. The mission was conceived and realised with the help from all ISRO Centres and payloads were realised by the payload PI Institutes in close collaboration with ISRO and many other national institutes - Indian Institute of Astrophysics (IIA); Inter-University Centre of Astronomy and Astrophysics (IUCAA); Laboratory for Electro-optics System (LEOS) of ISRO; Physical Research Laboratory (PRL); U R Rao Satellite Centre of ISRO; Vikram Sarabhai Space Centre (VSSC) of ISRO. We thank BARC-TIFR Pelletron facility at TIFR, Mumbai, India, and the staff therein for all the support during proton irradiation testing. Especially Dr. J. P. Nairh, Dr. H. Sparrowh, Dr. R. S. Worlikari, Dr. Anit Guptah, and Dr. Sanjoy Sen are acknowledged for their support. We thank the facilities and staff of U.R. Rao Satellite Centre, ISRO, for their support during different qualification tests. We acknowledge special assistance and guidance from Dr Kinshuk Gupta, URSC, ISRO. SP acknowledges the Manipal Centre for Natural Sciences, Centre of Excellence, and Manipal Academy of Higher Education (MAHE) for facilities and support during the development of the Thermal Filter Assembly.

\subsection*{Disclosures}
The authors declare there are no financial interests, commercial affiliations, or other potential conflicts of interest that have influenced the objectivity of this research or the writing of this paper.

\bibliography{v2_main}
\bibliographystyle{spiejour} 
\listoffigures

\vspace{1ex}

\end{spacing}
\end{document}